\documentclass[letterpaper, 10 pt, conference]{ieeeconf}  

\IEEEoverridecommandlockouts                              

\usepackage[pdftex]{graphicx}
\graphicspath{{figs/}}
\DeclareGraphicsExtensions{.pdf,.jpeg,.png}

\usepackage{graphics} 
\usepackage{epsfig} 
\usepackage{mathptmx} 
\usepackage{times} 
\usepackage{amsmath} 
\usepackage{amssymb}  

\usepackage{subfigure}

\usepackage{url}
\usepackage[acronym]{glossaries}
\usepackage{multirow}
\usepackage{makecell}
\usepackage{placeins}
\usepackage{xcolor}
\usepackage{supertabular,booktabs}
\usepackage{tabularx}
\usepackage{balance}
\usepackage[colorinlistoftodos,prependcaption,textsize=tiny]{todonotes}
\newcolumntype{Y}{>{\centering\arraybackslash}X}

\title{\LARGE \bf
For the Thrill of it All: A bridge among Linux, Robot Operating System, Android and Unmanned Aerial Vehicles
}

\author{
        Daniel V. Ruiz, Leonardo A. Vidal,  and Eduardo Todt \\
        Department of Informatics, Federal University of Paran\'a (UFPR), Curitiba, PR, Brazil \\ 
         \textit{dvruiz@inf.ufpr.br, leonardo.vidal@escola.pr.gov.br, todt@inf.ufpr.br}
        } 

\begin{document}

\maketitle

\newacronym{uav}{UAV}{Unmanned Aerial Vehicle}
\newacronym{ros}{ROS}{Robot Operating System}
\newacronym{slam}{SLAM}{Simultaneous Localization and Mapping}

\thispagestyle{empty}
\pagestyle{empty}

\begin{abstract}

Civilian Unmanned Aerial Vehicles (UAVs) are becoming more accessible for domestic use. Currently, UAV manufacturer DJI dominates the market, and their drones have been used for a wide range of applications. Model lines such as the Phantom can be applied for autonomous navigation where Global Positioning System (GPS) signal are not reliable, with the aid of Simultaneous Localization and Mapping (SLAM), such as monocular Visual SLAM. In  this  work,  we  propose  a  bridge  among  different systems, such as  Linux,  Robot  Operating  System (ROS),  Android,  and  UAVs as an open source framework, where the gimbal camera recording can be streamed to a remote server, supporting the implementation of an autopilot. Finally, we present some experimental results showing the performance of the video streaming validating the framework.

\end{abstract}




\section{INTRODUCTION}

In recent years, there has been a growing interest in \glspl*{uav}, commonly referred to as drones, due to the increasing accessibility for domestic use. According to the Global Consumer Drone Market Report 2016-2020~\cite{market2016}, the  current top  vendors  are  DJI, Parrot, 3D  Robotics, Cheerson Hobby,  Walkera, and Yuneec. With currently the manufacturer SZ DJI Technology Co.(DJI), based in Shenzhen, China, being the global market share leader in drone aircraft sales and industrial use~\cite{bill_2015,skylogic2018}. According to SkyLogic Research~\cite{skylogic2018}, DJI is the dominant brand for drone aircraft purchases, with a 74\% global market share across all price points. With an even higher share (86\%) of the core  [\$1,000 , \$2,000] price segment.

Due to the accessible price, DJI \glspl*{uav} have been used for a diversity of works such as photogrammetry  to  analyze the accuracy of a digital elevation map~\cite{KRSAK2016276}, locating, identifying, and monitoring courtship and mating behavior of the green turtle~\cite{Bevan2016}, bridge related damage quantification~\cite{Ellenberg2016} and power line detection and tracking~\cite{Zhou2016}. 

To develop autonomous navigation for such DJI \glspl*{uav}, an interface with the hardware is necessary. To do so, the manufacturer offers a DJI Onboard Software Development Kit (SDK)~\cite{onboardsdk} and a DJI Mobile SDK~\cite{Mobilesdk}. The compatibility depends on the model. Jiang \emph{et al.}~\cite{framework2018} propose an
extendable flight system for commercial UAVs that employ the \gls*{ros}~\cite{Quigley09}. However, they focus on the Onboard SDK available for a specific model line, the Matrice, while lines such as the most popular Phantom are not supported.

According to Jiang \emph{et al.}~\cite{framework2018}, \gls*{ros} has been largely used to build and test robot applications. One of the most attractive features of ROS is its distributed and modular design. 
In this case, each algorithm can be made into an independent module without several difficulties, such as
drivers of different hardware components and communications between multiple threads. Furthermore, most of the
state-of-the-art aerial robotics groups ~\cite{Bloesch2017,Leutenegger2015,Yang2017,Liu2017,Lin2017} have implemented their algorithms on \gls*{ros}.

In this work, we propose an open source framework that is essentially a bridge for the DJI Mobile SDK to be used together with the \gls*{ros} environment, allowing the use of \gls*{slam} algorithms already \gls*{ros} compatible such as ORB-SLAM2~\cite{orb2} to achieve autonomous navigation.

\section{RELATED WORK}

In this section, DJI Phantom and Matrice drones related works are briefly presented. Jiang \emph{et al.}~\cite{framework2018} developed an extendable flight system for commercial \glspl*{uav}, and they focus on the DJI Onboard SDK with the \gls*{ros} framework. Sagitov \emph{et al.}~\cite{sagitov2017} propose a more realistic simulation of the DJI Phantom 4 using the \gls*{ros} framework and Gazebo robot simulator~\cite{Koenig2004}. 

Z. Lu \emph{et al.}~\cite{Lu2017} propose four basic functionalities, obtaining compass information, controlling a gimbal, autopilot function for return, and video preview, which are developed using the DJI Mobile SDK and are implemented for iOS devices.

Sa \emph{et al.}~\cite{SaKKPNS17} present full dynamics system identification using only onboard Inertial Measurement Unit (IMU) that is used by a subsequent model predictive controller (MPC) based position controller, their system relies on the DJI Onboard SDK and focus on the DJI Matrice line.

Y. Lu \emph{et al.}~\cite{smart2017} developed a mobile application for Android 6.0 platform on Huawei MT7-CL00 as the system server. The Google Glass and Moto360 smart devices make the flight and camera control of DJI Phantom 3 Professional \gls*{uav} with Android Mobile SDK providing their interfaces. The wearable devices’ sample frequency is set at 50Hz. The aim of this project is an implementation using commercial devices.

\begin{figure*}[t!]
\centering
\includegraphics[width=0.9\textwidth]{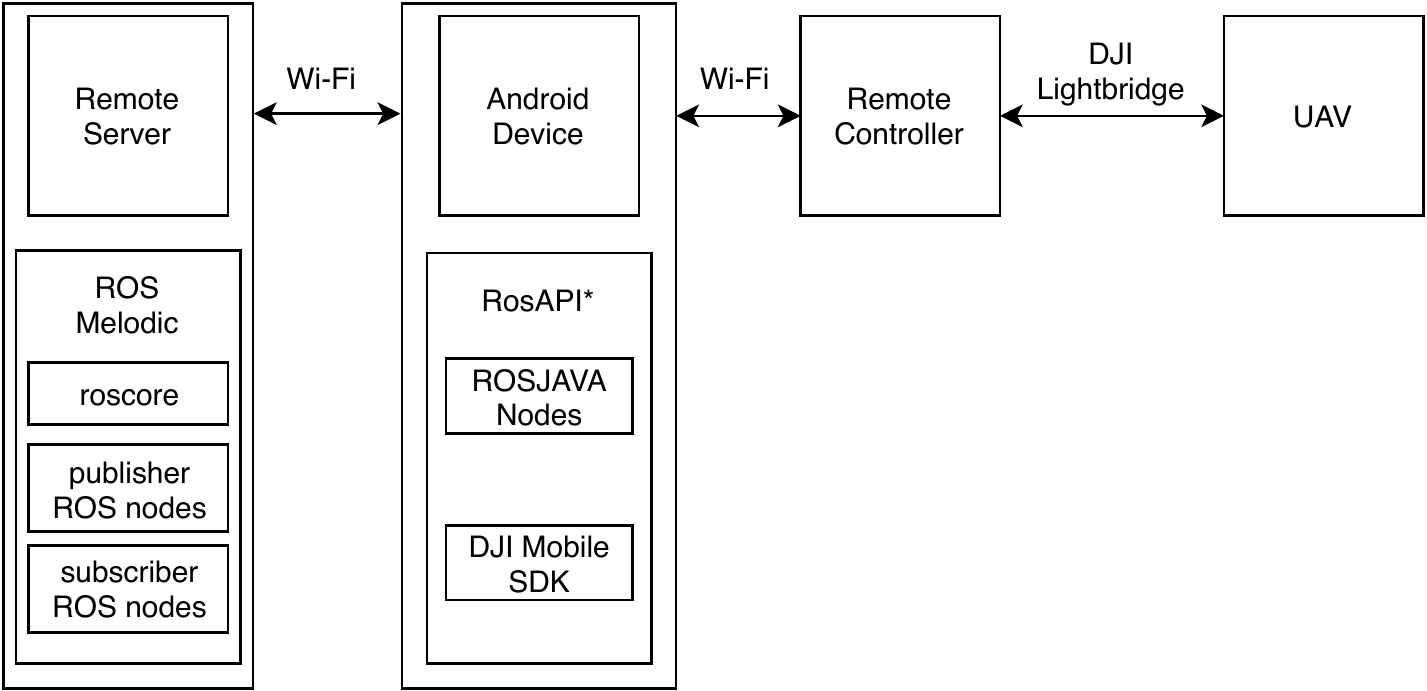}
\caption{Flowchart of operation. ROS messages are sent and received between the Android Device and the Remote Server via Wi-fi. The Mobile SDK manages the messages between the remote controller and the Android device. Communication between the UAV and the remote controller are managed by their firmware. RosAPI* is used for the proposed Android application.}
\label{fig:app}
\end{figure*}

Christansen \emph{et al.}~\cite{Christiansen2017} design and test a \gls*{uav} mapping system that uses a LiDAR sensor which can map the overflown environment in point clouds. LiDAR data are combined
with data from the Global Navigation Satellite System (GNSS) and IMU sensors to conduct environment mapping for point clouds.  The LiDAR point clouds are recorded, mapped, and analyzed using the functionalities of the \gls*{ros} framework and the Point Cloud Library (PCL).

Landau and van Delden~\cite{Landau2017} describe an \gls{uav} system architecture implemented with DJI Mobile SDK in Swift programming language for iOS operating system. It enables a DJI Phantom 3 drone to be controlled through voice commands using Nuance speech recognition platform and regular expressions are used as language control.

Accuware offers a video streaming library~\cite{accuware}, called Dragonfly DJI streamer library, which only supports Android 7 or above. Its license is proprietary and its source code is not publicly available. The library uses the WebRTC~\cite{webrtc}, an Application Programming Interface~(API) for browser based Real-Time Communications~(RTC).

The related work here presented have some limitations; for instance, the proposed autonomous navigation systems only operate with the Onboard SDK, which is for the Matrice line only. The systems compatible with the mobile SDK are limited by the smartphone utilized and usually limited by GPS signal or to a different mode of Human-computer interaction such as voice and gesture, instead of a fully autonomous system.

\section{PROPOSED WORK}

\begin{figure}[h!]
\centering

\subfigure[X, Y, and Z Axes]{\includegraphics[width=0.4\textwidth]{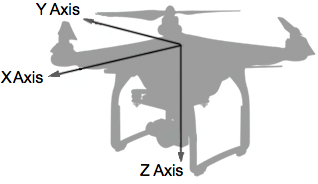}}
\vfil
\subfigure[Roll, Pitch,  and Yaw Axes]{\includegraphics[width=0.4\textwidth]{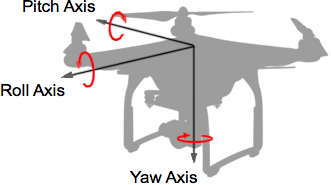}}

\caption{Body Frame coordinate system employed by navigation. Source: DJI~\cite{flight}}
\label{fig:body}
\end{figure}

\begin{figure*}[t!]
\centering
\includegraphics[width=0.9\textwidth]{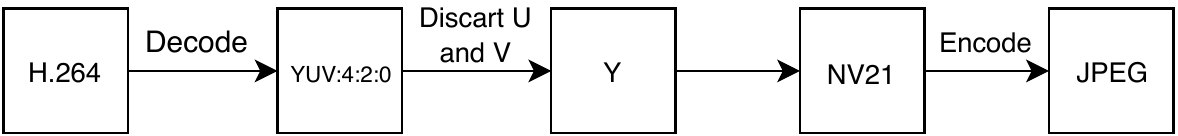}
\caption{Image format conversions. H.264 as video streaming input and JPEG as image frame output. A optimization trick is used to convert the YUV:4:2:0 format to NV21 using grayscale, by removing the UV channels (chrominance) there is no need to change the UV planes to an interleaved format.}
\label{fig:color}
\end{figure*}

The Mobile SDK, available for iOS and Android, has broader compatibility, supporting the Mavic, Spark, Inspire, Matrice, and Phantom series, while the Onboard SDK is only compatible with the Matrice line. We propose a link between technologies, in order to shorten this gap and enable a computer with higher computational power, running an operating system, such as Linux, to operate the \gls*{uav}, so the goal is an Android application that works as a bridge and allows the use of ROS nodes to control the \gls*{uav}.

There are places where Global Positioning System (GPS) signal is not reliable, such as indoors, canyons and close to tall buildings, hindering flight missions which heavily rely on those coordinates for navigation. To mitigate this \gls{slam} can be used, in this work, we focus on video streaming so that each frame can be used as input for a Visual \gls{slam} algorithm.

Fig.~\ref{fig:app} present a flowchart of the execution of the proposed system. The Android application, here  named RosAPI, allows the heavy processing to be done on a remote server,  with the use of the rosjava, an implementation of \gls*{ros} in pure Java with Android support.

The navigation uses the body frame coordinate system and operates using the virtual stick control from the Mobile SDK. Commands are sent via ROS nodes, which should be sent in a frequency of between 5 Hz and 25 Hz, are interpreted by the SDK in the Android and sent to the remote controller and to the drone. The gimbal motor can be operated via messages in the publisher-subscriber model employed as well. See Fig.~\ref{fig:body}. Flags for skip frame and compression rate are also available.

Image streaming from the gimbal camera is encoded as H.264~\cite{wiegand2003}, and the decoder employed in the mobile SDK, which is based on the FFmpeg library, decode it to YUV:4:2:0. The goal is to stream grayscale images due to their smaller size in bytes, thus speeding up transmission over the Wi-Fi. To further compress the byte size, an intermediary NV21 image format is used and then encoded as JPEG~\cite{wallace1991}.  An optimization trick was employed to convert the YUV:4:2:0 format to the intermediary NV21 image format and obtain an image in grayscale.  The trick consists of separating the Y channel (luminance) and discarding the UV channels (chrominance), thus mitigating the computational cost of changing the UV planes to an interleaved format. Conversion to NV21 is necessary due to the native YuvImage class from the Android (API), responsible for converting YUV images to JPEG, which requires NV21 encoding. See Fig.~\ref{fig:color}.

\section{EXPERIMENTS}

The experiments performed consist of an alternating image between white and black every two seconds. The \gls*{uav} camera record this event and feed forward to the remote controller which send to the android device and the remote server. The remote server is responsible for controlling and displaying the alternating image and receiving the perceived frame. Thus the $\delta_d$ difference between the timestamp of the exhibition of the image and the timestamp of the first frame that display that event can be used as the system delay. A timer is used for measurement of the $\delta_f$ between each received frame, which is then used for the Frame Per Second (FPS) metric. Five different smartphones were evaluated, see Table~\ref{tab:devices} for more detailed description of each device. The experiments were performed with a DJI Phantom 3 Standard drone.

\begin{table}[ht]
\centering
\caption{List of Android Devices Evaluated} 
\resizebox{0.48\textwidth}{!}{
\begin{tabular}{cccc}
\toprule
\textbf{Manufacturer} & \textbf{Device}    &  \textbf{CPU}         &  \textbf{GPU}          \\ \midrule
\multirow{3}{*}{Samsung}    & Galaxy J2 Prime \vspace{0.2cm}           & Quad-core 1.4 GHz Cortex-A53 & Mali-T720MP2                \\

                            & \multirow{2}{*}{Galaxy A8} & 2x2.2 GHz Cortex-A73 \&      & \multirow{2}{*}{Mali-G71}   \\
                            &                            & 6x1.6 GHz Cortex-A53         &                             \\ \midrule
ASUS                        & ZenFone Go                 & Quad-core 1.0 GHz Cortex-A53 & Adreno 306                  \\ \midrule
\multirow{2}{*}{Xiaomi}     & \multirow{2}{*}{Mi Mix 3}  & 4x2.8 GHz Kryo 385 Gold \&   & \multirow{2}{*}{Adreno 630} \\
                            &                            & 4x1.7 GHz Kryo 385 Silver    &                             \\ \midrule
Motorola                    & Moto G6                    & Octa-core 1.8 GHz Cortex-A53 & Adreno 506                  \\ \bottomrule
\end{tabular}
}
\label{tab:devices}
\end{table}

The quality flag used for the JPEG encoding was $90$ for all devices, and a skipframe was set to $2$. The experiment was conducted in an indoor environment were multiple Wi-Fi networks were operating. The obtained results are displayed in Table~\ref{tab:tests}.

\begin{table}[ht]
\centering
\caption{Video streaming experiment. FPS stands for Frames per Second. Delay is measured in miliseconds.} 
\resizebox{0.48\textwidth}{!}{
\begin{tabular}{cccccc}

\toprule
\multicolumn{1}{c}{\textbf{Device}} &
\multicolumn{1}{c}{\textbf{Android}} &
\multicolumn{2}{c}{\textbf{FPS}} &
\multicolumn{2}{c}{\textbf{Delay}} 
\\ \midrule

\makecell{} &
\makecell{} &
\makecell{Avg} &
\makecell{Std} & 
\makecell{Avg} &
\makecell{Std} 

\\ \midrule

\makecell{\textbf{Galaxy J2 Prime} \\ \textbf{Galaxy A8} \\ \textbf{ZenFone Go} \\ \textbf{Mi Mix 3} \\ \textbf{Moto G6}  } &
\makecell{\textbf{6.0.1} \\ \textbf{9} \\ \textbf{6.0.1} \\ \textbf{9} \\ \textbf{9}} &
\makecell{13.04  \\ 12.50 \\ 9.57 \\ 13.49 \\ 8.20} &
\makecell{0.48   \\ 2.60 \\ 1.47 \\ 4.57 \\ 2.10} &
\makecell{319.90 \\ 409.20 \\ 1295.95 \\ 545.12\\ 1280.70} &
\makecell{154.09 \\ 241.82 \\ 551.18 \\  371.02\\ 826.91}

\\ \bottomrule

\end{tabular}
}

\label{tab:tests}
\end{table}

The data show that the Galaxy J2 Prime presents the most stable streaming, the Mi Mix 3 with the highest FPS, while the MOTO G6 had the worst performance both in FPS and Delay. Since VSLAM algorithms rely on visual information the highest FPS is desirable, but due to the multi-hardware wireless communication nature of the system, there are physical limitations on the throughput of the frames, specially since decoding is done on a smartphone. Problems due to delay can be mitigated with the help of the IMU measurements. Higher delays implies in lower flying speeds to keep reliable navigation. There are still room for optimization on the framework aiming to obtain higher FPS and lower delay.

\section{CONCLUSIONS}

In this work, we propose a bridge among different systems Linux, Robot Operating
System, Android, and Unmanned Aerial Vehicles. Bridging the gap of using DJI Mobile SDK with ROS can motivate future works using the Phantom series with an autonomous flight robust enough to operate without a reliable GPS signal. Furthermore, it was left for future works the analysis of decoding the H.264 stream in the remote server, instead of doing it in the Android device, this could be a potential improvement on the transmission since H.264 supports Hardware-accelerated decoding on GPU, which could provide a more consistent frame rate and lower delay for the frames.

Additionally, our proposed framework ease the addition of other modules such as image processing, classification, object detection, semantic segmentation, and some others novel deep learning methods that explore domain adaptation and data generation that can run on the remote server and make use of Hardware-accelerated Deep Neural Networks running on GPU~\cite{zhang2019detecting,krinski2019mask,ruiz2019anda,ruiz2020giraffes}.

\bibliographystyle{IEEEtran}
\bibliography{main}

\end{document}